\documentclass{pasj01}
\Received{$\langle$reception date$\rangle$}
\Accepted{$\langle$acception date$\rangle$}
\Published{$\langle$publication date$\rangle$}

\usepackage{natbib, aas_macros}
\usepackage{bm}

\usepackage{ulem}

\begin{document}

\title{A Possible Time-Delayed Brightening of the Sgr A* Accretion Flow
after the Pericenter Passage of G2 Cloud}
 \author{Tomohisa \textsc{Kawashima} \altaffilmark{1,2,*},
 Yosuke \textsc{Matsumoto} \altaffilmark{3,4},
  and Ryoji \textsc{Matsumoto} \altaffilmark{4}}
\altaffiltext{1}{Center for Computational Astrophysics, National
 Astronomical Observatory of Japan, Mitaka, Tokyo 181-8588, Japan}
\altaffiltext{2}{Division of Theoretical Astronomy, National
 Astronomical Observatory of Japan, Mitaka, Tokyo 181-8588, Japan}
 \altaffiltext{3}{Institute for Global Prominent Research,
Chiba University, 1-33 Yayoi-cho, Inage-ku, Chiba 263-8522, Japan}
\altaffiltext{4}{Department of Physics, Graduate School of Science,
Chiba University, 1-33 Yayoi-cho, Inage-ku, Chiba 263-8522, Japan}
\email{kawashima.tomohisa@nao.ac.jp}

\maketitle

\begin{abstract}
A possibility of time-delayed radio brightenings of Sgr A*
    triggered by the 
pericenter passage of the G2 cloud is studied by carrying
out global three-dimensional magnetohydrodynamic simulations
taking into account the radiative cooling of the tidal debris
    of the G2 cloud.
Magnetic fields in the accretion flow are
strongly perturbed and re-organized after the passage of G2.
We have found that the magnetic energy in the accretion flow
increases by a factor 3-4 in 5-10 years after the pericenter
passage of G2 by a dynamo mechanism driven by the
magneto-rotational instability. Since this B-field
amplification enhances the synchrotron emission from the
disk and the outflow, the radio and the infrared luminosity of Sgr A*
is expected to increase around A.D. 2020. The time-delay of the
radio brightening enables us to determine the rotation
axis of the preexisting disk.
\end{abstract}

\section{Introduction}
The pericenter passage of an object named G2 close to the galactic
center black hole (BH) Sgr A* \citep{2012Natur.481...51G} was expected to
enhance the activity of Sgr A* by supplying the tidally stripped gas to
the BH.
The distance of its pericenter from the Galactic center BH is
only
${\sim} 2\times10^3r_{\rm s}$
\citep{2013ApJ...763...78G,2013ApJ...774...44G}, where $r_{\rm s}$ is the
 Schwarzschild radius.
Br-$\gamma$ observations indicate that the size of the G2 is ${\sim} 15 $ mas, i.e.
$ {\sim} 10^3r_{\rm s}$, which is as large as its pericenter distance.
The estimated mass of the gas component of G2 is 
$\sim$3$M_{\oplus}$, which is comparable to that of the Sgr
A* accretion flow, i.e., a hot accretion flow onto the Galactic center BH 
\citep[for hot accretion flows, see][and references therein]
{2008bhad.book.....K,2014ARA&A..52..529Y}. 
The G2 was, therefore, expected to affect the dynamics of
the Sgr A* accretion flow and trigger a flare.
However, no brightening event induced by the pericenter
approach of G2 has been observed in Sgr A*
\citep{2015ApJ...798L...6T,2015ApJ...802...69B}. 

After the discovery of the G2, a number of numerical simulations have
been performed   
\citep{2012ApJ...750...58B,2012ApJ...755..155S,2013MNRAS.433.2165S,2014MNRAS.440.1125A,2014PASJ...66....1S}. 
Most of these simulations did not take into account magnetic fields
despite their important roles in accretion processes
\citep[e.g.,][]{1991ApJ...376..214B,1995ApJ...446..741B,1999ASSL..240..195M,2000ApJ...528..462H,2012MNRAS.423.3083M}.
\cite{2013MNRAS.433.2165S} carried out
 three-dimensional (3D) general relativistic magnetohydrodynamic (MHD)
 simulations 
in order to study the shock structure at which the electrons may be
accelerated, during the pericenter passage of the G2 cloud.
Long-term MHD simulations of 
the G2 cloud, however, have not yet been carried out,
 although it is
necessary to explore the future brightening of Sgr A*.

In this paper, we present the possibility of the
time-delayed amplification of magnetic fields and subsequent accretion
by carrying out longer time-scale 3D MHD simulations of a
hot accretion flow impacted by a cloud.
Since the gas supplied by the cloud need a several rotation period at the
pericenter to settle into a rotating torus, and typically 10 rotation period
for the re-organization of magnetic fields, it takes 5-10 years until
the magnetic fields are re-amplified by the dynamo action driven by the
magneto-rotational instability (MRI).
Therefore, in order to study the possibility of the time delayed
brightening, we need to carry out MHD simulations for
time scales of a decade.
This is the motivation of this paper.

It is still controversial whether or not G2 harbors a star in its
center: \cite{2015ApJ...798..111P} observed G2 in infrared band using SINFONI
and proposed that G2 is a pure gas clump, which is possibly formed from
a gas stream around Sgr A*, while \cite{2014ApJ...796L...8W} observed G2
in L’ band using Keck observatory and proposed that G2 harbors a star
because the size of the dust of G2 does not change during its pericenter
passage.
For simplicity, we assume pure gas clouds in this paper.
 This assumption should be reasonable, partly because a
tidally disrupted gas feature is evident in $p$-$v$
diagram \citep{2015ApJ...798..111P}, and partly because the size of the
region where gravity 
of the central star dominates that of the Galactic center BH is only
${\sim} 1$\% of the estimated cloud size \citep{2014ApJ...796L...8W}, so that
we can  neglect the effect of the central star even if it exists.

\section{Simulation Model}

We solve a set of MHD equations taking into
account the anomalous resistivity \citep{1994ApJ...436L.197Y} and the 
effects of radiative cooling in cylindrical coordinates $({\varpi},
{\varphi}, z)$.
In this paper, ${\varpi}$ and $r$ denotes the cylindrical and the
spherical radius, respectively.
The equations are the same as those in \cite{2006PASJ...58..193M},
except we incorporate the effects of radiative cooling due to
recombination with approximate cooling function \citep{2014ApJ...786L..12G}:
\begin{equation}
\Lambda_{\rm rec} =
10^{-21} (\rho/m_{\rm p})^2 \exp[-(9/2)(\log T - 4)^2] ~ {\rm erg} ~
{\rm cm}^{-3} ~  {\rm s}^{-1},
\end{equation}
as well as the bremsstrahlung emission $\Lambda_{{\rm brem}} = 6.2 \times 10^{20}
\rho^2 T^{1/2} ~  {\rm erg} ~ {\rm cm}^{-3} ~  {\rm s}^{-1}$, since the 
structure of the G2 cloud should be  
affected by the radiative cooling
\citep{2014ApJ...786L..12G,2014PASJ...66....1S}.
Here, $m_{\rm p}$, $\rho$, and $T$ is the proton mass, mass density, and
gas temperature, respectively.
In this work, we do not incorporate the effects of the synchrotron
cooling.
 It is well known that, in optically-thin hot accretion disks,
 the bremsstrahlung cooling dominates the synchrotron cooling in the
 outer disk, which we simulate in this work, while the
 synchrotron cooling can be dominant in the inner disk.
 This is because the synchrotron cooling rate is roughly proportional to
 $T^{1}$ if we assume the energy equipartition,
 while the bremsstrahlung cooling rate is proportional to $T^{1/2}$.
Since the gas temperature decreases with radius, the
 bremsstrahlung emission dominates the synchrotron radiation in the
 outer disks.
In this work, as we mention below, we calculate the outer accretion
flow, in which the bremsstrahlung emission is the dominant radiative process.
In addition, the G2 cloud is so cool that synchrotron emission cannot
dominate the Br-$\gamma$ and bremsstrahlung emissions, since the
synchrotron photons are radiated by relativistic electrons.
These are the reasons why we ignored the synchrotron cooling in this work.
The anomalous resistivity is incorporated by using the same formula as
that employed in \cite{2006PASJ...58..193M}, i.e., $\eta = \eta_0 [{\rm
max}((j/\rho)/v_c -1, 0)]^2 $, where $\eta$ is the magnetic diffusivity,
and $j$ is the current density. 
There are two free parameters: the coefficient for the anomalous
resistivity $\eta_0$ and the threshold value $v_c$.
We set $\eta_0 = 1.25 \times 10^{-3}$ and $v_{\rm c} \sim 69.5$ in our
simulation unit (i.e., $v_{\rm c} = 0.9c$). 

We carry out global 3D simulations by using a newly-developed
MHD code CANS+ \citep{2016arXiv161101775M}. 
The code is based on the HLLD approximate Riemann solver proposed by
\cite{2005JCoPh.208..315M}.
In order to preserve monotonicity and to achieve high-order accuracy in
space, we employ a monotonicity 
preserving, fifth-order accurate interface value reconstruction method, MP5
\citep{1997JCoPh.136...83S}. 
A third-order TVD Runge-Kutta method is used for the time integration. 
We adopt the hyperbolic divergence cleaning method
\citep{2002JCoPh.175..645D} in order to minimize numerical errors of the
divergence free condition of the magnetic field.

The number of computational cells is $(N_{\varpi}, ~N_{\varphi}, ~N_z) = (256,
~128, ~320)$.
The cell size is constant in the inner region: ${\Delta}{\varpi}  =
{\Delta}z  = 30 r_{\rm s}$ for
$0 < {\varpi} < {\varpi}_0$  and  $|z| < {\varpi}_0$.
 Here ${\varpi}_0$
is the location of the pressure maximum of the initial torus, which
we set $\varpi_0 ~ \equiv ~ 3{\times}10^3 r_{\rm s}$.
Outside this region, ${\Delta}{\varpi}$ and ${\Delta}z$ increase with
${\varpi}$ and $z$, respectively: ${\Delta}{\varpi}_{n} = {\rm
min}(1.05{\Delta}{\varpi}_{n-1}, ~{\Delta}{\varpi}_{\rm max})$,
${\Delta}z_{n} = {\rm min}(1.05{\Delta}z_{n-1}, ~{\Delta}z_{\rm max})$
for $z>0$, and  
${\Delta}z_{n} = {\rm min}(1.05{\Delta}z_{n+1}, ~{\Delta}z_{\rm max})$
for $z<0$. 
Here, $n$ denotes the sequential cell number, and 
${\Delta}{\varpi}_{\rm max}$ and ${\Delta}z_{\rm max}$ are set to
be $200r_{\rm s}$.
In azimuthal direction, the cell size is set to be constant, i.e.,
${\Delta}{\varphi} = (2{\pi}/N_{\varphi})$.
The computational domain is, thus, $0 \le {\varpi} \le
4.323{\times}10^{4}r_{\rm s}$, 
$0 \le {\varphi} \le 2{\pi}$, and $|z| {\le} 2.778{\times}10^4 r_{\rm s}$.
A spherical absorbing inner boundary is imposed at $r_{\rm in} =
450r_{\rm s}$, i.e., the physical quantity $q$ is approximated by 
$q = q^{\prime} - c^{\prime}(q^{\prime} - q_0)$ if their cell-centers are inside 
$r_{\rm in}$, where $c^{\prime}$ is the damping coefficient, $q^{\prime}$ is 
the numerically obtained quantity at the next timestep, and $q_0$  is the initial value 
[for details, see equation (7) and (8) in \cite{2006PASJ...58..193M}].
The shape of this absorbing boundary modestly matches up with the
exact spherical one, since the cell size is $\sim$ 5\% of the inner boundary radius.
The outer boundaries are the free boundaries where waves can be
transmitted.

At first, we perform the simulations of hot accretion flows without
introducing a gas cloud until the accretion flow attains
a quasi-steady state. 
After the quasi-steady state is realized, a gas cloud in pressure
equilibrium with the ambient gas is located into the
simulation box.

\subsection{A Hot Accretion Flow Model for the Sgr A*}

We assume a rotating, equilibrium torus with the pressure maximum at
${\varpi} = {\varpi}_0$. 
The torus is assumed to be threaded by a weak, purely toroidal, initial
magnetic field, by using the equilibrium solution of magnetized tori
proposed by  
\cite{1989PASJ...41..133O}.
The plasma $\beta$ at the pressure maximum of the torus is initially 100.
We set the direction of the rotation axis of the initial torus to be the same as
$z$-axis.
The torus is embedded in a hot, isothermal, non-rotating, static, coronal
atmosphere with gas temperature ${\sim} 1.3{\times}10^{10}$ K. 
For more details of the simulation set-up, see, e.g.,
\cite{2006PASJ...58..193M}.

\begin{figure*}[!th]
\begin{center}
  \includegraphics[scale=0.555]{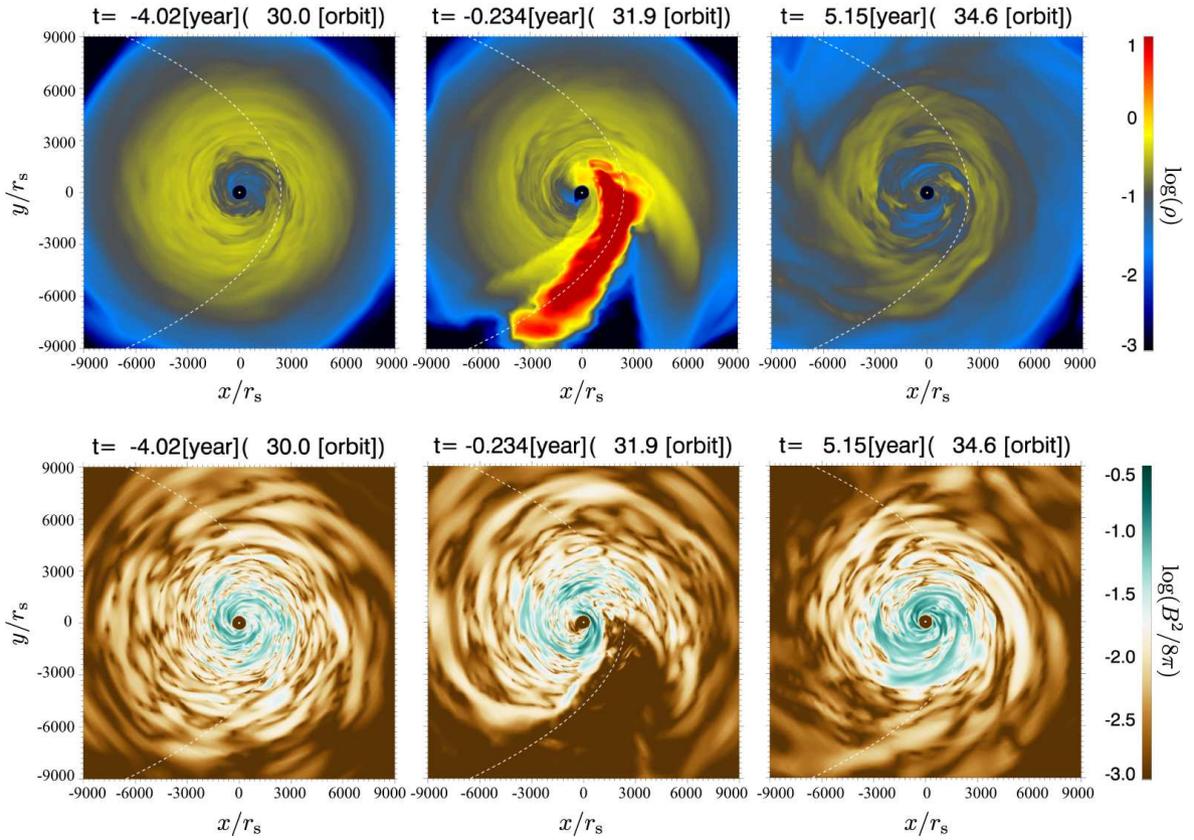}
 \end{center}
  \caption{Snapshots of the simulated accretion flow before and after
 the passage of the gas cloud for the
 model with $i = 0$.
 Color maps show the mass density (top) and the magnetic energy density
 (bottom).
 The pericenter passage time estimated by using test particle
 approximation is defined as $t=0$ yr.
 The dashed white curves represent the Kepler orbit of the center of G2.
 } 
 \label{fig:snapshots_0eq}
 \end{figure*}

After the growth of the non-axisymmetric mode of MRI ($\sim 10$
orbital-period at the pressure maximum of the initial torus), the
angular momentum is efficiently transported by the Maxwell
stress and an accretion flow is formed.
At 30 orbital-time at the pressure maximum, we locate a gas cloud in the
computational  
domain to simulate the interaction of the G2 cloud with the Sgr A*
accretion flow.

The normalization factor of the mass density 
${\rho}_0$ is chosen to be consistent with the observational implication of mass
density distribution of the Sgr A* accretion flow
${\rho}_{\rm RIAF}(r) = 1.3{\times}10^{-21}(r/10^{4}r_{\rm
s})^{-1.125}~{\rm g} ~ {\rm cm}^{-3}$
\citep{2003ApJ...598..301Y,2012ApJ...759..132A} such that 
${\rho}_{0}{\hat \rho (\varpi_0)}$ $\sim$ ${\rho}_{\rm
RIAF}(\varpi_0)$, 
where ${\hat \rho}(\varpi_0) \sim 0.3$ is the azimuthally averaged, normalized
mass density of the quasi-steady accretion flow.

\subsection{Models of the G2 Cloud}

There are six parameters describing orbits of a point mass:
the orbital inclination $i$, the longitude of the ascending node $\Omega$,
the argument of the pericenter $\omega$, the eccentricity $e$, the
semi-major axis $a$, and the time of the pericenter passage $t_0$.
The three parameters $e$, $a$, and $t_0$ can be constrained from the
observations of the G2. 
According to the
observation of the Br-$\gamma$ emission line and the data analysis by
\cite{2013ApJ...774...44G}, we 
set $t_0 = {\rm A.D.}~ 2014.25$, $e=0.9762$ and the 
pericenter radius $r_{\rm p} = a(1-e)=2.4{\times}10^3 r_{\rm s}$.
 The semi-major axis $a$ is obtained from $e$ and $r_{\rm p}$.
For convenience, we replace $t-t_0$ by $t$ so that the G2 passes
the pericenter at $t=0$ yr.

For our simulations, the other three parameters $i$, $\Omega$, and
$\omega$ cannot be constrained from the observation, because of the
uncertainty of the angle between the rotation axis of the Sgr A*
accretion flow and that of the Galactic plane. 
For simplicity, we assume ${\Omega} = 0$ and ${\omega} = 0$, where
${\omega} = 0$ means that the pericenter is assumed to be on the
equatorial plane of the accretion flows.
We expect that the parameter $\Omega$ does not significantly affect the results,
because the global structure of the accretion flow is not highly
non-axisymmetric.
In this paper, we present results for $i=0$ and
${\pi}/3$ rad.
In this work, we assume a Schwarzschild BH and employ 
a pseudo-Newtonian potential \citep{1980A&A....88...23P}, so that we do
not take into account the relation between the direction of the
BH spin and the orbit parameters of the gas cloud described above.

We set the initial position of the center of G2 at
$r=2.4{\times}10^{4}r_{\rm s}$. 
For the initial cloud density, we assume 
Gaussian distribution with FWHM $=
3{\times}10^{15}$ cm \citep{2013MNRAS.433.2165S} in such a way that
the total mass of the gas cloud is 3$M_{\oplus}$.
The initial velocity inside the cloud is assumed to be equal to the
Kepler orbital velocity at its center of mass.  
For the sake of simplicity, 
we do not assume the initial magnetic field in the cloud.
We note that ${\bm {\nabla}}{\cdot}{\bm B} = 0$ is assured when we locate the
cloud, because the magnetic field is neither artificially added to nor
removed from the computational domain.

The cloud which satisfy the assumption above is located in the computational domain
after a quasi-steady accretion flow is formed.

\section{Results}
Time evolution of the gas cloud and the hot accretion flow when $i=0$ is
shown in figure \ref{fig:snapshots_0eq}.
At $t~\simeq -4$ yr (i.e., 30 Kepler orbital time at the initial pressure
maximum of the torus), nonlinear growth of non-axisymmetric MRI has already
been saturated, so that the accretion flow has attained the
quasi-steady state. 
At this time, the spherical gas cloud is located at
$2.4{\times}10^4r_{\rm s}$, far outside the zoomed region of figure \ref{fig:snapshots_0eq}.
At $t~\simeq 0$ yr, the gas cloud stretched by the tidal force of the
Galactic center BH
 penetrates the accretion flow.
This stage is qualitatively the same as \cite{2013MNRAS.433.2165S},
except that the tidally stretched gas becomes slimmer 
due to the effects 
of the radiative cooling in this work.
At $t~{\simeq 5}$ yr, the accretion flow returns back to the
quasi-steady state with its magnetic field being amplified by the
MRI-driven-dynamo compared to
that before the passage of the gas cloud.
The mass density and pressure at $r \gtrsim 3\times 10^3 r_{\rm s}$
is roughly half of those before the G2 encounter because
the disk mass is swept by the G2 impact. 
However, the variation of mass density and gas pressure inside $2 \times 10^3 r_{\rm s}$ (i.e., inside the pericenter radius of the G2 cloud) is about several tens percent, while the magnetic energy increases by a factor 3-4 after the pericenter passage.
Therefore, we expect the radio brightening of the Sgr A* by
 the synchrotron emission
 when the gas with the B-field, which is  amplified via the MRI-driven-dynamo
 triggered by the G2 encounter, accretes onto the innermost region
 of the accretion flow with a time-delay, i.e., the dynamo timescale
 and the accretion timescale at $\sim 10^3 r_{\rm s}$.
It takes not only dynamo timescale but also accretion timescale to show the brightening, because the synchrotron emission from the innermost region of the pre-existing accretion flow is too bright to mask the radio brightening due to the enhanced magnetic field at $\sim 10^3 r_{\rm s}$. 
When the gas with amplified B-field at $10^3 r_{\rm s}$ accretes to the innermost region, synchrotron emission is expected to increase.

\begin{figure}[!h]
\begin{center}
  \includegraphics[scale=0.27]{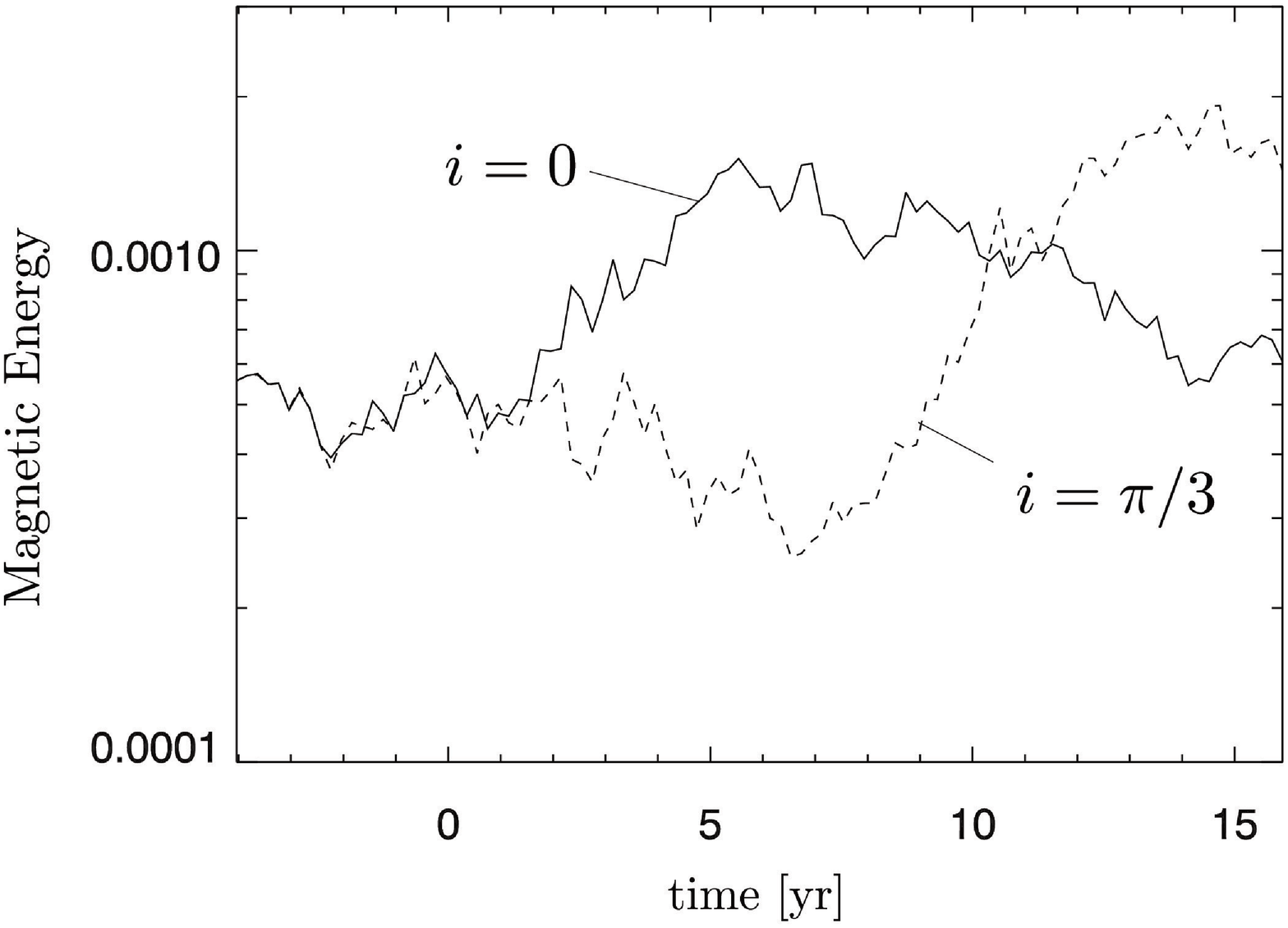}
 \end{center}
 \caption{Time evolution of magnetic energy obtained by spatial
 integration of the magnetic energy density inside $750r_{\rm
 s}$.
 The solid and dashed
 curves represent the results for the models with 
 $i=0$ and $\pi /3$, respectively.
The magnetic energy is normalized by $(\rho_0 v_0^2/2)\varpi_0^3
 \simeq 1.47 \times 10^{44}$ erg,
 where $\rho_0
 \simeq 1.68 \times 10^{-20}$
 ${\rm g}~{\rm cm}^{-3}$ 
 and $v_0 \simeq 3.87 \times 10^{8}$ ${\rm cm}~{\rm s}^{-1}$ are the
 initial torus mass density and 
 the Keplerian velocity at $\varpi = \varpi_0 = 3 \times 10^3 r_{\rm s}$,
  respectively.
   } 
 \label{fig:evolution_energy}
 \end{figure}

\begin{figure}[!h]
\begin{center}
 \includegraphics[scale=0.3]{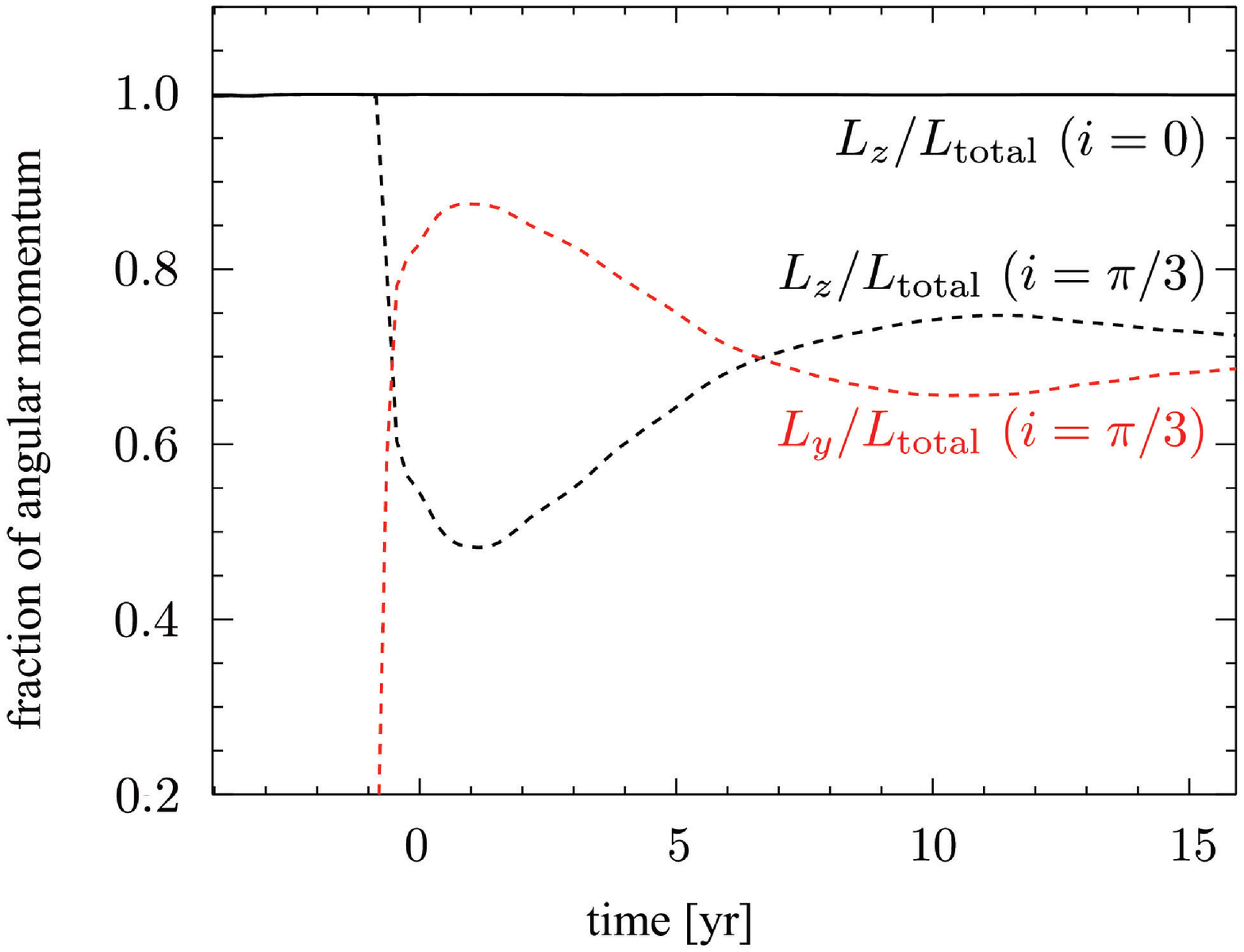}
 \end{center}
 \caption{Time evolution of angular momentum integrated inside $\varpi =
 750r_{\rm s}$. Each component of angular momentum is normalized by the
 total angular 
 momentum $L_{\rm total} = \sqrt{\mathstrut L_x^2 + L_y^2 + L_z^2}$. 
   } 
 \label{fig:evolution_L}
 \end{figure}

 In figure \ref{fig:evolution_energy}, we present time evolution of
magnetic energy. 
 The peak magnetic energy, which is 3-4 times larger than that before
 the encounter with G2, appears about 5 and 13 yrs after the pericenter
 passage of the G2 cloud for the model with inclination angle $i = 0$ and $\pi/3$,
 respectively.
In the case that $i=0$, magnetic energy increases after
the impact of the G2 cloud because the
cloud collision increases the radial component
of magnetic fields, which is subsequently
amplified by differential rotation of the disk.
In the case that $i=\pi/3$, magnetic energy slightly
decreases after the cloud impact because the
enhanced turbulent motion dissipates the magnetic
energy. When the rotation axis of the disrupted
disk is fixed, MRI-driven-dynamo
amplifies the magnetic field of the tilted disk.

Here, let us discuss a little more about the competition between the
magnetic dissipation and amplification.
  Since magnetic turbulence enhanced by the passage of the gas cloud
  increases the 
magnetic dissipation, the accretion flow approaches to a Taylor state
\citep{1974PhRvL..33.1139T}, in which the system relaxes to a state
with minimum magnetic energy. Figure \ref{fig:evolution_energy}
indicates that the relaxation of the magnetic energy is more significant
in the model with $i=\pi/3$, in which the system is highly perturbed by
the impact of the cloud.
Subsequently, magnetic field is amplified by MRI through the generation
of $B_z$ by turbulence.
Numerical results for $i=\pi/3$ indicates that the magnetic field amplification
by the latter mechanism becomes dominant 5 yrs after the pericenter
passage of G2.
This time scale is consistent with that of the disk
dynamo at $\varpi \sim 10^3r_{\rm s}$.

\begin{figure*}[h]
\begin{center}
 \includegraphics[scale=0.68]{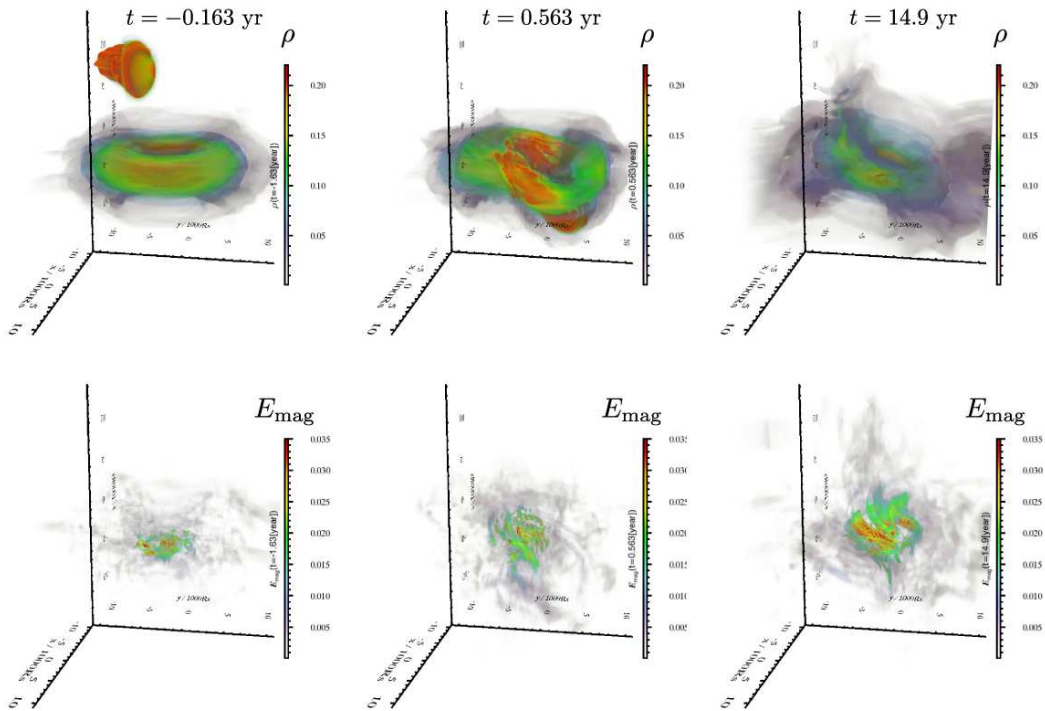}
 \end{center}
 \caption{Volume rendered image of the simulated accretion flow with the gas
 cloud for the model with $i = \pi/3$.
 The top and bottom panels show the distribution of mass density and
 magnetic energy density, respectively.  
 \label{fig:snapshots_60}
 } 
 \end{figure*}

As mentioned above, the B-field
 amplification is delayed in 
 the model with $i = \pi/3$ because it takes time before the decay of
 strong perturbation caused by the impact of G2.
 Figure \ref{fig:evolution_L} shows time evolution of the angular
 momentum of the flows integrated such as ${\bm L} = \int_{\varpi_{\rm in}}^{750r_{\rm
 s}} \int_0^{2\pi} \int_{-\varpi_0}^{\varpi_0}\rho({\bm r}\times {\bm v})~
 \varpi d\varpi d\varphi dz $.
For the model with $i=\pi/3$, the direction of the
 angular momentum is remarkably modified after the G2 passage:
 At $t \sim 0$ yr, the accretion flow is strongly disturbed by 
 the cloud impact (see also figure \ref{fig:snapshots_60}) and the
 direction of angular momentum drastically changes.
 After $t~\sim~5$ yr, the fluctuation of the $L_y$ and $L_{\rm z}$ is less
 than ${\simeq}$ 10\%, so that it can be regarded that the flow has settled
 to a quasi-steady state with a tilted rotation axis. 
 In this quasi-steady disk, the B-field
 amplification by MRI restarts to 
 dominates the decay of the magnetic field.
 Thus, the amplification of the B-field in the model with $i=\pi/3$
 is delayed by 5 yrs.

  In figure \ref{fig:snapshots_60}, it is shown that the accretion flow is
 tilted after the
 passage of the gas cloud for the model with $i=\pi /3$.
 The tilt of the accretion flow is caused by 
 the angular momentum transport from the gas cloud to the accretion flow,
 because the angular momentum of the accretion flow is misaligned with
 that of the initial orbital motion of the gas cloud.
 The angular momentum of
 the gas cloud is sufficient to change the direction of the angular
 momentum  vector of the accretion flow.
 Since it requires several years for the rotation axis of the accretion
 flow to settle into the new direction, the B-field amplification via
 the disk dynamo is delayed when the inclination is large.

\section{Summary and Discussion}

We carried out 3D MHD simulations of the interactions of the Sgr A*
accretion flow with the gas cloud G2, taking into account
the effects of radiative cooling.
We found that the magnetic energy increases by 3-4 times in 5-10 years
after the pericenter passage of the G2 cloud.
The delay time of the B-field amplification depends on the orbital inclination
of the gas cloud: the maximum magnetic energy appears $\simeq$5 and
$\simeq$13 yrs after the pericenter passage for the model with $i = 0$ and
$\pi/3$, respectively. 
The B-field amplification can increase the radio and the infrared
luminosity 
with a time delay after the G2 passage.
We expect that the gradual increase of the synchrotron emission with a peak
around A.D. 2020 can be observed in the radio and the infrared bands.
This significant radio brightening should occur when the amplified
magnetic field accretes to the innermost region.
Furthermore, the X-ray flare may occur when the amplified
magnetic energy is released via the
magnetic reconnection in the vicinity of the BH. 

Here, we discuss the consistency of our scenario with the no detection
of the increased radiative flux in Sgr A* up to today.
In this paper, we have found that the magnetic
field in the Sgr A* accretion flow is amplified 5-10 yrs after the pericenter
passage of the G2 cloud.
Our scenario would reasonably explain the no detection of the radio and
the infrared
brightening
to date, because the radio and the infrared emission
in the Sgr A* may be dominated by the synchrotron emission, which should
be enhanced with the B-field amplification.

By contrast, the X-ray emission of Sgr A* may be dominated by the
bremsstrahlung emission from the outer part of the accretion flow at
${\sim} 10^5 r_{\rm s}$ \citep{2002ApJ...575..855Q,2003ApJ...598..301Y}.
This radius is far outside the distance of the
pericenter of G2.
Thus, change of the dynamics of the inner accretion flow 
induced by the G2 impact will not affect the X-ray luminosity of Sgr
A*, except the  
X-ray flare induced by the magnetic reconnection in the vicinity of the
BH.
It may be thought that the G2 should affect the X-ray luminosity when the
cloud starts to interact with the accretion flow at $10^5r_{\rm s}$.
However, since the size of G2 is only ${\sim 10^3 r_{\rm s}}$, it would be
too small to affect the dynamics of the accretion flow at $10^5 r_{\rm s}$.
No detection of increased X-ray emission up to today is, therefore, 
consistent with our 
scenario.

If the brightening is not detected during the next ${\sim}~10$ yrs,
there can be two possible reasons: the location of the outer edge of the
accretion flow is closer to the Galactic center BH than that of
pericenter of th G2, or the gas component of G2 would be
less massive than the expected one.
The expectation for the brightening discussed above should be confirmed
by calculating time-dependent 
multi-wavelength radiative spectra, by post-processing the MHD simulation data.
In subsequent papers, we would like to carry out the spectral calculations,
as well as 
the parameter survey of the MHD simulations of the disk-cloud
interaction.

At the beginning of our simulations, the
Br-$\gamma$ luminosity obtained by volume integration of the cooling
function expressed in equation (1) attains the luminosity $\sim 10^{-3}
L_{\odot}$, which is consistent with the observed luminosity of the G2
\citep{2012Natur.481...51G}. 
During the pericenter passage, however, the Br-$\gamma$ luminosity decreases
down to $\sim 10^{-4} L_{\odot}$  which is one order of magnitude
lower than the observed one \citep{2015ApJ...798..111P}.
This inconsistency may be caused by overheating of the G2 cloud due to
the numerical mixture of the G2 cloud and the hot accretion flows
during the pericenter passage.
This problem would be solved by performing the simulations with 
higher spatial resolution, as shown in simulations with adopted mesh
refinement (AMR) code focusing not on the system including both the Sgr A* 
accretion flow and G2 cloud but on only G2 cloud \citep{2015ApJ...811..155S}.
However, the B-field amplification shown in this work should occur also in 
simulations with fine spatial resolution, since it is driven by the
disk dynamo (especially by MRI) and the high-resolution-simulations
rather show more efficient B-field amplification \citep{2016Sci...351.1427H}.
Simulations with higher spatial resolution reproducing the
consistent luminosity of the Br-$\gamma$ emission remains as a future work.

It should be noted that, after the pericenter passage of the G2
cloud, the mass accretion rate at the inner boundary ($450 r_{\rm s}$) increases 2-4 times
of that before the G2 impact in our simulations. 
However, since the
magnetic energy inside the disk does not increase
until the MRI driven dynamo grows again (i.e.,
5-10 years after the G2 impact), the synchrotron
luminosity would not increase significantly until
the strongly magnetized region begins to infall.

Wnen the gas with the amplified B-fields accrete onto the
inner disk, synchrotron emission from the inner
disk will increase.
Although the observational images would not be perfectly the same as those
predicted by \cite{2012ApJ...752L...1M} because of the amplification of the magnetic field after the G2 encounter in our study,
synchrotron brightened region may be similar to that studied by
\cite{2012ApJ...752L...1M}.
The brightening may be detected by East Asia mm/submm VLBI
observation and Event Horizon
Telescope submillimeter Very Long Baseline Interferometry experiment
(EHT).
As discussed below, the angle between the rotation axis of the accretion
disk and the orbital axis of the G2 
may be constrained by these observations.

Let us discuss whether or not the direction of the
rotation axis of the preexisting accretion flow can be constrained by
the timing of the brightening.
The radio brightening in the vicinity of the BH
is expected to follow the amplification of the B-field at ${\sim} ~ 10^3
r_{\rm s}$, i.e., 
the increased B-field will be advected inward and, subsequently, will be
amplified further near the BH.
If we assume that the amplified B-field is advected to the
innermost region of the accretion flow in the viscous accretion timescale, the
time lag due to the advection is 
${\lesssim} ~1$ yr, where we have assumed viscosity parameter
\citep{1973A&A....24..337S} $\alpha \simeq 0.1$ since
our simulations indicate this value in the B-filed re-amplification stage.
This accretion timescale is shorter than the timescale of the B-field
amplification ${\sim}$5 or ${\sim}$10 yrs so that we can identify 
the difference of the orbital inclination of G2 against the
preexisting accretion flow.
The comparison of the timing of brightening in the simulation
with the future observation would enable us to constrain the 
direction of the rotation axis of the preexisting Sgr A* accretion flow,
since the orbital plane of the G2 is known. 
Furthermore, the tilt of the accretion flow in
$i=\pi/3$ case (figure 4), can be spatially resolved by the
East Asia mm/submm VLBI observation and EHT.
If the direction of the rotation axis of the accretion disk significantly differs from the angular momentum axis of G2, we would be able to observe the change of the rotation axis of the accretion flow with time.

The tilt of the accretion flows caused by the disk-cloud interaction
can occur not only in Sgr A* but also in the other low luminosity active
galactic nuclei (LLAGNs).  
The tilt may induce the quasi-periodic oscillation in the LLAGNs and/or
the change of the direction of the LLAGN jets.
These possible behavior would be important to explore the accretion and
ejection histories in LLAGNs.

In this work, we set the inclination $i=0$ and $\pi/3$, i.e., the gas cloud is assumed to be co-rotating with the accretion disk. 
If the gas cloud is counter-rotating with the accretion disk,  we expect that the retrograde gas cloud will lose more angular momentum than the prograde one does affected because of the ram pressure of the accretion flow.
Especially in the perfectly retrograde case
(i.e., $i = \pi$), the most part of gas cloud may not be able to keep the Keplerian orbit, which is
estimated by the observations of
Br-$\gamma$ emission line,  until the G2 reaches 
the pericenter. 
It may also excite a strong
disturbance of the accretion flow due to the mixing of the gas with opposite angular
momentum, which would result in the drastic increase of mass accretion rate onto the black hole.
We leave the parameter study of the gas cloud including the
counter-rotating case as a future work.

 \begin{ack}
  We thank Y. Feng, K. Ohsuga, H.R. Takahashi, M. Kino, and M. Akiyama for
   useful discussion. 
The numerical simulations were mainly carried out on the XC30 at the Center for
  Computational Astrophysics, National Astronomical Observatory of
   Japan.
   This research also used computational resources of the HPCI system provided
by the Information Technology Center, the University of Tokyo,
and Research Institute for Information Technology, Kyushu University
through the HPCI System Research Project (Project ID:hp120193, hp140170).
This work was supported in part by MEXT HPCI
STRATEGIC PROGRAM and the Center for the Promotion
   of Integrated Sciences (CPIS) of Sokendai, 
   and MEXT as a priority issue
   (Elucidation of the fundamental laws and evolution of the universe)
   to be tackled by using post-K Computer and JICFuS.
This work was also supported by JSPS KAKENHI Grant Number 16H03954, and the NINS project of Formation of International Scientific Base and Network.
 \end{ack}


\end{document}